# Personalization of learning using adaptive technologies and augmented reality

Maiia V. Marienko[0000-0002-8087-962X], Yulia H. Nosenko[0000-0002-9149-8208] and Mariya P. Shyshkina[0000-0001-5569-2700]

Institute of Information Technologies and Learning Tools of the NAES of Ukraine, 9 M. Berlynskoho Str., Kyiv, 04060, Ukraine
`{popel, nosenko, shyshkina}@iitlt.gov.ua`

**Abstract.** The research is aimed at developing the recommendations for educators on using adaptive technologies and augmented reality in personalized learning implementation. The latest educational technologies related to learning personalization and the adaptation of its content to the individual needs of students and group work are considered. The current state of research is described, the trends of development are determined. Due to a detailed analysis of scientific works, a retrospective of the development of adaptive and, in particular, cloud-oriented systems is shown. The preconditions of their appearance and development, the main scientific ideas that contributed to this are analyzed. The analysis showed that the scientists point to four possible types of semantic interaction of augmented reality and adaptive technologies. The adaptive cloud-based educational systems design is considered as the promising trend of research. It was determined that adaptability can be manifested in one or a combination of several aspects: content, evaluation and consistency. The cloud technology is taken as a platform for integrating adaptive learning with augmented reality as the effective modern tools to personalize learning. The prospects of the adaptive cloud-based systems design in the context of teachers training are evaluated. The essence and place of assistive technologies in adaptive learning systems design are defined. It is shown that augmented reality can be successfully applied in inclusive education. The ways of combining adaptive systems and augmented reality tools to support the process of teachers training are considered. The recommendations on the use of adaptive cloud-based systems in teacher education are given.

## 1 Introduction

Nowadays, augmented reality is being used in education to improve students' learning facilities as well as the quality of learning [18].

The Industrial Revolution 4.0 has changed how production processes are organized and economic development goals are achieved, and this has affected the education sector, and has given rise to the term Education 4.0. As noted in [1, p. 23] Education 4.0 consists of the personalized learning process, game-based learning using Virtual Reality/Augmented Reality (VR/AR), communities of practice, adaptive technologies, learning analytics, and E-Assessment. This has led to unlocking the potential of AR technology use as a component of students' adaptive learning environments [29; 30]. The use of AR can help students to learn better, improve learning settings and increase the quality of the educational process.

The tools of adaptive technologies and VR/AR make the educational process more interactive, actualize the cognitive interest of students, and enhance motivation for learning [7; 27]. In this case, VR/AR allows you to interact with real and virtual objects, vividly and realistically visualize didactic material [15]. Most current courses using this technology follow a "one size fits all" approach that addresses the common needs of different people. Instead, adaptive technologies are geared toward meeting the needs of each student, as they can get adjusted, adapt properly, depending on user behavior [39]. In our opinion, it is a promising trend of research for education to introduce teaching methods that will combine the adaptive technologies with VR/AR, as this will allow to achieve better personalization of the educational process.

## 2 Literature review

Numerous domestic and foreign scientists have addressed the study of various aspects of the creation and use of adaptive cloud-oriented systems. Vladimir G. Sragovich emphasizes that it is not necessary to simultaneously manage and assess the entity [37]. That is, the adaptive system changes its algorithm (or its structure) automatically, which assumes that the goal is achieved under all conditions. Elena V. Kasianova [16] believes that networked learning systems combine intellectual learning systems and adaptive media systems. Besides, according to Elena V. Kasianova's research, all adaptive hypermedia systems can be combined into one class, which can be considered as hypertext and hypermedia systems. Due to this, each user will have a workplace adapted with individual tools and customization of various aspects of the system itself (without affecting the work of other users). Aleksander A. Gagarin and Sergiy V. Tytenko considered adaptability in systems of continuous learning [11]. Therefore, in their work adaptability is considered first of all as a symbiosis of purpose and result. At the same time, the goal is considered by the one that the user of the system puts forward for himself, and as the result is the educational result obtained by the user at this stage of work with the system.

In the framework of the conducted research Pavlo I. Fedoruk stated that by using adaptive and intellectual technologies, the educational system allows taking into account the student's abilities, his/her previous knowledge, skills [9].

In particular, Svitlana H. Lytvynova considered the design of a cloud-oriented environment in a general secondary education institution [19] that has the features of an adaptive system; Pavlo I. Fedoruk studied the methodology of organizing the process of individualized learning using an adaptive system of distance learning and knowledge control; Yurii V. Tryus considered pedagogical mentoring as an element of adaptive management in the system of preparation of future teachers [40].

Mariya P. Shyshkina in [36] considered the possibility of combining cloud technologies and adaptive systems while emphasizing that several models of cloud services can be used simultaneously: SaaS and PaaS. Mykola V. Pikulyak (2016) [34] proposed to build an adaptive training module in the distance education system based on the script method. The proposed idea is based on scenario examples as a separate special rule (software solution) that binds quanta (units of material).

Gary Natriello outline even more complex learning environments when discussing the possibilities of transformative VR games [26]. They rely on four types of meaningful engagement as a way to think about adaptive educational technologies that can support transactive engagement with adaptive systems. They compare the four types of content interaction as:

1. Procedural training,
2. Conceptual training or understanding how the tools work,
3. Consistent training or learning about the impact of their actions on the contexts developed,
4. Reflects the impact of their actions on the developed context.

In the study of National Academy of Education [25] the content of the concept of adaptive technologies examined, the potential of their use in education analyzed, the infrastructure required for successful implementation is characterized, etc.

Luis de-la-Fuente-Valentín, Aurora Carrasco, Kinga Konya and Daniel Burgos reveal predictions about what technologies will be used shortly in education [7]. The latest forecasts are the introduction of new concepts such as 3D printing [14], the Internet of Things [13], learning analytics [31], massive open online courses [32] and AR [38].

Lisa Balme in [2] presents the results of a survey of teachers on their experience in using adaptive technologies working with students with disabilities, their vision of the benefits of these technologies, and the obstacles to implementation, and more.

Philip Kerr describes the basic concepts, the use of adaptive technologies in English language learning, teacher training, and development opportunities are presented [17].

Moses Basitere and Eunice Ivala consider the effectiveness of the use of the adaptive Wiley Plus ORION training platform in the study of physics has been analyzed, the results of testing using the adaptive test and the standard blank (paper) test have been compared in [3].

Rosliza Hasan, Faieza Abdul Aziz, Hesham Ahmed Abdul Mutaleb and Zakaria Umar in [12] used many mobile technologies – from simple recording devices and audience response systems to VR/AR apps for tablets. The paper presents an architecture built to organize existing digital interactive web content for learning through the concept of modules, submodules, and essentials.

Yigal Rosen, Ilia Rushkin, Rob Rubin, Liberty Munson, Andrew Ang, Gregory Weber, Glenn Lopez and Dustin Tingley in [35] describes the results of an experiment on implementing adaptive functionality in a mass open online course (MOOC) based on edX. The advantages and disadvantages of this technology that need refinement are indicated.

Hayatunnufus Ahmad, Norziha Megat Mohd Zainuddin and Rasimah Che Mohd Yusoff considered the use of AR technology to better memorize Quran for students with hearing impairments. The research is aimed at developing an integrated mobile application that can facilitate the memorization of the Quran among hearing-impaired students [1].

Arwen H. DeCostanza, Amar R. Marathe, Addison Bohannon, A. William Evans, Edward T. Palazzolo, Jason S. Metcalfe and Kaleb McDowell in [6] suggest the development of the mechanisms to increase team effectiveness, in particular military teams, for heterogeneous teams using technologies focused on improving teamwork through individualized information, processes and activities for each team member. In particular, scientists have explored the use of VR/AR technology.

Synaptic Global Learning, in collaboration with the Center for Innovation and Excellence in eLearning at the University of Massachusetts (USA), developed the world's first adaptive MOOC in computational molecular dynamics, called aMOOC, based on Amazon Web Services cloud architecture [3].

The mobile technologies have been used to support the learning process in research of Yevhenii O. Modlo et al. – from audience response systems to virtual reality and augmented reality applications for mobile Internet devices [20; 21; 22; 23; 24]. The paper [1] presents an architecture built to organize existing digital interactive web content for learning through the concept of modules, submodules and essentials, among which are separate widgets.

Thus, adaptive learning technologies are based on the use of the most relevant and up-to-date student data, and collaborative teams are formed instantly on the request of teachers. This leads to the development of adaptive hypermedia systems and personalization of learning experiences and the creation of a personal learning environment in higher education. Collective and cooperative learning is no exception. Researchers look at a variety of strategies for group tasks within a similar environment, dividing students into groups, exploring the creation of dynamic groups and different models of group work in a pre-planned scenario, leading to improved learning [26].

Most research in the field of adaptive learning environments have been concerned to creating customized solutions designed to digitize a specific part of the curriculum used by teachers [33]. Such approaches lead to limited use of the system, typically requiring a significant amount of work and resources to be expanded with new techniques such as adaptive, collaborative and cooperative learning, unless they initially focus on their direct support. This is mainly because the introduction of support for new techniques in existing, advanced learning technologies usually means significant changes to existing databases of training systems.

Therefore, cloud-oriented approaches are a promising way of developing adaptive educational systems. Flexible and open, they are more focused on improving the learning environment, the introduction of new components, supplying computing power based on user needs, customization of personal information processing and information needs of the user.

Despite numerous partial studies of special issues of adaptive learning systems and AR technologies application, the methods of these technologies' educational use remain relevant and poorly understood subject matter.

Particularly relevant is the problem of developing theoretical and methodological foundations for the use of adaptive cloud-based systems in combination with AR technologies. This is the key to the training of competent ICT professionals, highly qualified teaching staff for modern education.

*The purpose of this research is* to consider the state of the art and the approach for adaptive cloud-based systems design and to develop the recommendations for educators on using adaptive technologies and augmented reality in personalized learning implementation.

## 3 The personification of learning as a leading global educational trend

Turning to the theory of adaptive systems, the task is to build a controller that will affect a specific object and over time will (under any circumstances) achieve the goal. The system consisting of the object parameters and the specified controller will be called adaptive [10]. The time to reach the goal is called the time of adaptation.

Due to [8] the inherent adaptability of the service can be manifested in one or more aspects: content, evaluation, and sequence (fig. 1).

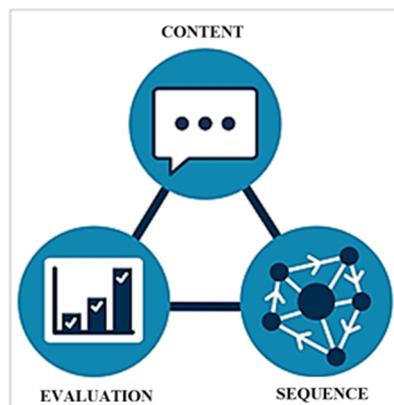

**Fig. 1.** Aspects of service adaptability [8].

Services with adaptive content allow you to determine what kind of educational material the student doesn't understand. The system "splits" each training block into parts, and the student can move to the next content unit only after successfully mastering the previous ones. If you have problems with a specific content block, the system monitors and prompts you to re-go through the content on that topic. At the same time, the educator can track the progress of each student: at what pace he/she is completing tasks, and so on. An example of a platform with adaptive content is CK-12 [5]. This is a free English language resource with educational materials in most school subjects. The typical content of a training block in CK-12 is video, textual examples, interactive assignments (PLIX), simulations [5].

Adaptive assessment services are designed in such a way that each question/task depends on how the student handled the previous one: if successful, the following question/task offered is more complex, if unsuccessful, then simpler. Such services can be created as part of certain applications, platforms (ST Math, Smart Sparrow, etc.), and separately (Typeform, Quizalize, etc.) [28]. Adaptive tests can be selected and applied depending on didactic goals – as a means of ongoing, thematic and/or final assessment, monitoring. For example, MAP Growth service is recommended to be used for periodic assessment of various subject matters (a "long" test every few months, and MAP Skills is recommended to be used more frequently since it allows you to determine what difficulties a student has when mastering the material and to properly adjust the learning process [28]. An important advantage of monitoring using adaptive tests is the ability to obtain detailed statistics for each student at different intervals – from months to years, and accordingly, build an individual learning trajectory (alone or with the help of an educator) [28].

Adaptive sequence services collect and analyze user data continuously. While the student is working on tasks, the program analyzes the answers and sequentially selects the next content of the corresponding complexity. In general, such programs can take into account different indicators: the correctness of the answers to the questions, the number of attempts, the time spent, the use of additional resources, personal interests (for example, which resources the student prefers), and sometimes mood [8]. The development of such services is the most time consuming since they allow us to build and adjust the individual educational trajectory of each student in real-time. The adaptive sequence goes through three steps: data collection, data analysis, adaptation of the material flow to the needs of each student [8]. The adaptive sequence is applied by Knewton (https://www.knewton.com/).

Sometimes developers of adaptive services apply several adaptive aspects. For example, the Precalculus online course combines adaptive assessment and adaptive sequencing (ALEKS service). Smart Sparrow platform has adaptive content and sequence (https://www.smartsparrow.com/what-is-adaptive-learning/), etc.

The world's leading companies and institutions are investing in advanced digital technologies, such as mobile communications, online social media, big data analytics systems, "intelligent" devices that control connected objects and sensors, and more. Hybrid cloud solutions emerge as a promising direction for technological development and implementation of the latest educational systems [36].

The development of adaptive educational systems, mostly with elements of artificial intelligence, requires the processing of large amounts of knowledge obtained from students. Due to cloud services that implement high-speed computing, the possibility of dynamic adaptation to the achieved level of knowledge, experience, skills of the learner is achieved. Therefore, with the use of hybrid cloud solutions, educational systems are becoming more adaptable, based on the integration of different types of services and their integration into teacher education into a single environment [36].

In this regard, a number of important trends can be identified that characterize the promising avenues for the development and use of modern learning personification technologies:

— The personification of learning that is achieved on the basis of "intellectualization" of all links of educational systems, their further integration in the educational process and the learning environment;
— The development of adaptive cloud-oriented platforms, their further unification, universalization, the formation of common standards for the development and implementation of individual modules, subsystems and training systems with elements of artificial intelligence;
— The growing role of the Big Data approach for collecting and analytics on learning outcomes and individual student progress;
— The development of information and analytical tools of the educational and scientific environment in the direction of their greater "intellectualization", use of advanced methods of semantic and syntactic analysis of data and texts in the process of searching for necessary information, processing of requests provided in natural language, use of elements of AR when working with text;
— The increasing saturation of the learning environment with a variety of intelligent devices, remote controls, robots, peripheral equipment, etc. that can be managed on a single platform, over the network (Internet of Things);
— The development and implementation of systems of educational and scientific cooperation in virtual teams with the use of augmented (virtual) agents, including technologies of individualization and enhanced collaboration between people and agents;
— The increasing role of computer literacy and technological culture for all participants in the learning process for the successful development and implementation of a new generation of artificial intelligence (AI) learning tools.

We anticipate advances in science and technology focused on personalized, adaptive technologies to improve teamwork, leading to new opportunities that revolutionize the way we train both individuals and teams to improve group collaboration. We also anticipate that this shift in focus to more basic and comprehensive incorporation of intelligent and adaptive technologies within the organization will lead to a change in critical knowledge, skills, and competences across different workplaces. Further, we expect that the study of shared mental models, such as transactive memory systems [6], will lead to the development of fully integrated human-agent (VR/AR) hybrid teams requiring new training using new techniques.

Some researchers consider assistive technologies as a separate subset of adaptive technologies. In a general sense, these are technologies which application provides support for specific activities for persons with special needs (SEN). Assistive technologies (ASTs) represent a wide range of tools, strategies, and services that meet an individual's individual needs, capabilities, and tasks, and include an assessment of an individual's needs with SEN, a functional assessment of the environment in which he/she resides, and selection, design, setting up, adapting, applying, maintaining, repairing and/or replacing services, coordinating them with educational and rehabilitation plans and programs for the comprehensive development and support of "education for all".

As a matter of fact, ASTs in educational systems are featured in adaptive technologies due to the ability to customize these systems to the needs of a wider range of users, e.g. persons with SEN. As a result of the pedagogically expedient implementation of the ASTs in the educational process gives students the opportunity to perform educational tasks with a greater degree of independence, with less effort.

Here are some examples:

— For students with visual impairments: Access Note, Aipoly Vision, VoiceOver, JAWS screen reader, the built-in Narrator system reader, NVDA reader, etc.;
— For students with hearing impairment: Fakih, AR-Book, ELRA, Vuforia
— For students with autism spectrum disorder: Expressionist, iCommunicate, Look2Learn, Proloquo2Go, etc.;
— For students with dyslexia and reading challenges: Audible, Learning Ally, Kami, Augmented Ally, Speechify, etc.

It is important to ensure that ASTs are accessible to vulnerable sections of the population. So, according to the World Health Organization, in low-income and middle-income countries, only 5-15% of people who require ASTs have access to them [41]. The availability of ASTs is greatly enhanced by the inclusion of VR/AR technologies. In this case, the user can use both single service and a range of services, depending on the needs and tasks.

Among the benefits for the students are the following:

— Involvement in the educational process and responsibility for one's own learning;
— Increased motivation and satisfaction with achievements;
— Positive attitude to the educational process, easier to perceive complex and abstract concepts;
— Development of practical skills;
— The use of different senses to receive and understand information.

## 4    Development and use of augmented reality technologies based on hybrid cloud solutions

The main difference between the new generation educational systems from the previous stages of the development of AI and computer-aided learning tools is a higher level of adaptability. It is achieved both through the use of more powerful and integrated student models and learning with AI elements, and the organization of a more flexible and open learning environment, in particular through hybrid cloud solutions, providing access to personalized services both individually and collectively [12].

By the *adaptive cloud-based learning system* the cloud-based system that has the property to be adjusted automatically by its parameters to the different individual characteristics and educational needs of the learning process participants is meant.

In order to implement the computer-procedural functions of this system, a virtualized computer-technological (corporate or hybrid) infrastructure should be purposefully created.

Thus, the most important characteristics of adaptive cloud-oriented educational systems, through which these systems have significant potential for use in pedagogical higher education systems, in particular pedagogical systems, include:

— availability of virtualized or hybrid ICT infrastructure;
— personalization of services;
— openness to modification and improvement;
— accessibility (use of open access, open data);
— the flexibility of algorithms for evaluating material complexity, student knowledge, readiness for learning, etc.;
— providing personalized assistance in the learning process;
— possibility of automatic adjustment on necessary parameters in real-time;
— systematic structure and functions.

VR training and assessment can have a progressive impact on productive learning [12]. VR systems are characterized as a human and computer environment in which users are integrated, ready to see, act, and collaborate with the three-dimensional world [12].

VR/AR has some learning benefits that are limited by traditional learning. For example, VR is ideal for dangerous workouts to avoid risk, allows you to present objects from different perspectives that are inaccessible under real conditions, allows you to visualize virtual equipment, allows you to use dynamic learning, gives students a sense of control as they can repeat the learning material as many times as they need and learn new topics at their own pace. A 3D interactive animation environment that attracts more attention than photography for students plays a positive role in learning. Besides, VR/AR can save time and money in the training of professionals, as they do not require the purchase of additional devices and provide training anywhere, anytime [12].

Starting and operating flexible manufacturing systems is extremely challenging because of the large number of output parameters. As tasks change over time, operators have no real way to learn or prepare to routinely solve their tasks. VR systems can fundamentally assist in operation and maintenance. Using VR, an administrator can get results online, and current directions can be guided by performing tasks that have never been done by downloading interactive media streams that could include motion, playback, the grouping of videos in real life, or their mixed set, from the so-called innovative multimedia base or even from a real-world framework (situation models). For example, a maintenance strategy is developed through an activity that demonstrates the development of the assembly and disassembly of the part, and the student controls the entire procedure [12].

The current problem is that all students who use VR have a shared training schedule that is not customized individually. However, each student learns at his/her own pace and focuses on specific parts of the task. Usually, given the end goal, adjusting or reworking the curriculum requires the human impact that is extremely large and painstaking. Computerization can stay out of this human influence [12].

Future reunions of VR/AR people and agents are giving rise to the idea of greater diversity and change in team and team members as you complete the training task. In particular, the anticipated benefits of these complex teams include combining the exact skills and skills required for a particular mission. Therefore, the ability to quickly bring together diverse teams of people and agents to work effectively in a group is needed. For example, with a long-term, ongoing knowledge of the strengths, status, and behavior of existing team members, and an individualized, adaptive agent may be responsible for working with team members to quickly assimilate individuals within the team, understanding the changing roles and responsibilities of team members, and facilitating shared understanding tasks and situations in the team [6]. VR/AR can offer a team during logical training, continuous monitoring of cognitive states, and communication to provide personalized, adaptive instruction when a team member seems to have misunderstandings or is out of sync with the group. That is, whether it is VR or AR, they contribute to the development of group work skills and their organization, interaction in a team.

## 5    Current research developments

From 2018 at the Institute of Information Technologies and Learning Tools of the National Academy of Educational Sciences of Ukraine (Ukraine), a planned scientific study "Adaptive cloud-based system of secondary school teachers' training and professional development" (2018-2020) is held.

In 2019 the results of the different aspects of the study were tested at 28 scientific and practical events: 6 conferences (4 international ones); 17 workshops (1 international). The problematic issues of scientific research were discussed and presented for the scientific and pedagogical community by organizing and conducting by the authors a series of training sessions, seminars, webinars for scientific and pedagogical staff, and graduate students.

In 2018 the V4+ Academic Research Consortium Integrating Databases, Robotics and Language Technologies was established, which aimed to address regional issues related to EU ICT research priorities: Partner search for Horizon 2020, building up digital platforms of the future, language barriers, technology-enhanced learning, scientific-cultural heritage, and know-how to exchange.

The main result of the V4+ACARDC project was the creation of a complex system of IT support consisting of the technological network infrastructure, educational software WPadV4 and a didactics methodology on how to create educational packages and associated learning materials and multi-lingual support. A cloud-based platform was used for sustainable information support of the project life cycle according to the jointly determined aims and information technological integration of the project management. The platform proved to be suitable to meet these needs and to perform smoothly and intuitively.

The functional model of the WPadV4 system was demonstrated and analysed in the previous publication [4].

The adaptive content management was supported by the research tool WPadV4, which was used to process available data in a sustainable model. Thus, all the data collected in the course of the research were findable, accessible, interoperable, and for all partners. This was to provide the openness and flexibility of the research collaboration processes.

*The learning platform* was considered as the set of the cloud-based tools to support different learning and research activities. Within the unite platform a lot of different tools may be integrated providing more opportunities to realize adaptive learning.

The question of choosing and integrating services, exploring their various components (adaptive technologies and VR/AR), as well as supporting open education and science systems, combining intelligent technologies and network services presents a prospect for further research that needs careful study.

As now the hybrid cloud-based solutions are at demand there is an important issue for further research is to consider and build different configurations in view of the basic principles and approach. The experimental design was based on several available services for the different components of the general model that was outlined. Still, the set of services is not still exhaustive in any case.

## 6       Recommendations and suggestions

With individualized, adaptive technologies and systems, VR/AR we can truly represent the revolutionary capacity for lifelong learning that can be incorporated into future concepts of uniting people and agents of virtual or augmented reality.

Combining adaptive cloud-oriented systems, VR/AR technologies and modern pedagogical techniques will be an effective solution to the problem, which will contribute to the adaptability of the education system to the individual characteristics of training pedagogical and scientific-pedagogical staff. The use of information technology, at this stage of educational development, is better focused on combining adaptive technologies with VR/AR technologies. These technologies are effective modern tools to personalize learning.

Adaptive VR can make learning more efficient and effective. AR learning can be improved by using adaptive, information-oriented models for adaptation and computerization of learning.

Conducting training and practice sessions in realistic learning environments or environments reminiscent of the future workplace increases the likelihood that professional competencies will help to increase productivity in the future, in the work process. There is already some experience in using realistic learning environments and modeling to work effectively with a team. Recent technological advances, including the proliferation and cost-effectiveness of VR/AR systems, combined with artificial intelligence to extend the experience, create the potential for large-scale implementation of the learning process in a realistic environment.

An important component of adaptive cloud-based learning systems is assistive technology. Developers and users of ASTs should follow the recommendations:

— Development/selection of ASTs according to the needs of a specific user group. ASTs should be compatible with tasks, emotional needs of users, their lifestyle, etc.;
— Support for the policy of providing vulnerable groups with access to the ASTs (point and strategic sponsorship support);
— Ensuring ease of use (availability of ASTs instruction manual that is accessible and understandable to any user who does not have) proper technical training;
— Involvement of users in the design of the ASTs - at different stages of development and testing.

Some ASTs include AR systems, which are particularly effective to support the Augmented Reality Operational hearing process [1].

Among the areas of adaptive cloud-based systems use for teachers training there are such as:

— To organize educational communication in a personalized mode, with the use of telecommunication tools, for example, the components of the public and the corporate cloud of the educational institution, as well as communication services and scientific and educational information networks;
— To support of individual and group forms of learning activities (classroom and extra-classroom) using the services of a scientific-educational cloud of an educational institution based on Microsoft Office 365, G Suite for Education, FaceTime, Google Duo, Meet and other;
— To use computer-based adaptive systems and platforms that were tested in different educational and socio-cultural environments and are now widely used in the world educational space: curriculum platforms (Alta, Cerego, Fishtree, Fulcrum Labs, LearnSmart, RedBird Advanced Learning, Smart Sparrow, Socrative); adaptive learning management systems, creation of training courses (Neo LMS, Open Learning Initiative (OLI)); adaptive testing systems (Typeform, Quizalize); adaptive adult learning platforms (Elevate) and more;
— To include cloud services of open science, in particular, services of European research infrastructures, in the composition of facilities and services of forming adaptive cloud-oriented systems in a pedagogical university; scientific and educational networks; cloud data collection, submission, and processing services; as well as the services of the European Open Science Cloud;
— To use adaptive content management tools based on a public cloud, for example, WPadV4 tool;
— To implement the methodology for supporting the adaptive knowledge-based processes of creation and use of e-learning resources and other kinds of services;
— To include the components of the corporate and public clouds of the educational institution (databases and data collections, adaptive content management systems, cloud-based office software applications, specialized software training tools, language processing tools, educational robots, and others) as well as services of publicly available information systems (scientific-educational information networks and infrastructures, cloud-based educational, scientific services) into teachers training;

— To provide visibility by constructing different interpretations of mathematical models, visualization of mathematical abstractions, etc. via AR tools;
— To provide accessibility through the use of a shared interface for access to environmental assets and reliable open source software; increasing temporal and spatial mobility;
— To form a unified learning environment, the content of which is developed in the learning process.

## 7    Conclusions

The main difference between the new generation educational systems from the previous stages of the development of computer-aided learning tools is a higher level of adaptability. It is achieved both through the use of more powerful and integrated student models and learning with artificial intelligence elements, and the organization of a more flexible and open learning environment, in particular through hybrid cloud solutions, providing access to personalized services both individually and collectively.

The development of adaptive educational systems, mostly with elements of artificial intelligence, requires the processing of large amounts of knowledge obtained from students. Due to cloud services that implement high-speed computing, the possibility of dynamic adaptation to the achieved level of knowledge, experience, skills of the learner is achieved. Therefore, with the use of hybrid cloud solutions, educational systems are becoming more adaptable, based on the integration of different types of services and their integration into teacher education into a single environment.

An important component of the adaptive cloud-based learning systems is assistive technology that cover a wide range of tools, strategies and services that meet the individual needs, capabilities and tasks of the individual, as well as the selection, design, setup, adapting, implementing, maintaining, repairing and / or replacing services, coordinating them with educational and rehabilitation plans and programs for comprehensive development of each individual.

Today a promising educational trend is to introduce teaching methods that will combine the adaptive technologies with VR/AR. Integration of adaptive cloud-oriented systems, the augmented reality technologies and the modern pedagogical techniques will be an effective solution to the problem of the adaptability of the education system to the individual features of educators' training, and will allow to achieve better personalization of the educational process. This issue needs further thorough research.